\documentclass[journal]{IEEEtran}
\usepackage[utf8]{inputenc}
\usepackage{graphicx}
\usepackage{hyperref}
\usepackage{enumitem}
\usepackage{geometry}
\usepackage[symbol]{footmisc}
\usepackage{subfigure}

\usepackage{cite}
\usepackage{amssymb, amsmath}
\usepackage{lineno}

\usepackage {dsfont}
\usepackage{algorithmic}
\usepackage{algorithm}
\usepackage{float}
\usepackage{times}
\usepackage{amssymb}
\usepackage{amsfonts}
\usepackage{makeidx}
\usepackage{color}
\usepackage{comment}
\usepackage{listings}
\usepackage{xcolor}

\definecolor{codegreen}{rgb}{0,0.6,0}
\definecolor{codegray}{rgb}{0.5,0.5,0.5}
\definecolor{codepurple}{rgb}{0.58,0,0.82}
\definecolor{backcolour}{rgb}{0.95,0.95,0.92}

\lstdefinestyle{mystyle}{
  backgroundcolor=\color{backcolour}, commentstyle=\color{codegreen},
  keywordstyle=\color{magenta},
  numberstyle=\tiny\color{codegray},
  stringstyle=\color{codepurple},
  basicstyle=\ttfamily\footnotesize,
  breakatwhitespace=false,         
  breaklines=true,                 
  captionpos=b,                    
  keepspaces=true,                 
  numbers=left,                    
  numbersep=5pt,                  
  showspaces=false,                
  showstringspaces=false,
  showtabs=false,                  
  tabsize=2
}

\begin{document}
\title{Large Language Models meet Network Slicing Management and Orchestration}

\author{\IEEEauthorblockN{Abdulhalim Dandoush\IEEEauthorrefmark{1},\IEEEauthorrefmark{2}, Viswanath Kumarskandpriya\IEEEauthorrefmark{1},  Mueen Uddin\IEEEauthorrefmark{2}, Usman Khalil\IEEEauthorrefmark{3}
}

\IEEEauthorblockA{\IEEEauthorrefmark{1}Esme Research Lab, SA ESME, Ivry-Sur-Seine, France} 
\IEEEauthorblockA{\IEEEauthorrefmark{2} University of Doha for Science and Technology (UDST), Doha, Qatar}
\IEEEauthorblockA{\IEEEauthorrefmark{3} University Brunei Darussalam, Brunei Darrussalam}
\IEEEauthorblockA{Corresponding author: Abdulhalim Dandoush, abdulhalim.dandoush@udst.edu.qa}
    }
\date{}

\maketitle

\begin{abstract}
Network slicing, a cornerstone technology for future networks, enables the creation of customized virtual networks on a shared physical infrastructure. This fosters innovation and agility by providing dedicated resources tailored to specific applications. However, current orchestration and management approaches face limitations in handling the complexity of new service demands within multi-administrative domain environments. This paper proposes a future vision for network slicing powered by Large Language Models (LLMs) and multi-agent systems, offering a framework that can be integrated with existing Management and Orchestration (MANO) frameworks. This framework leverages LLMs to translate user intent into technical requirements, map network functions to infrastructure, and manage the entire slice lifecycle, while multi-agent systems facilitate collaboration across different administrative domains. We also discuss the challenges associated with implementing this framework and potential solutions to mitigate them.
\end{abstract}

\section{Introduction}
The ever-growing demand for diverse network services with varying Quality-of-Service (QoS) requirements has pushed the boundaries of traditional network architectures. Network slicing emerges as a transformative technology addressing this challenge \cite{DAN21}. It allows the creation of isolated virtual networks (VNs) on top of a shared physical infrastructure. These slices can be customized with dedicated resources (connectivity, storage, computing) to cater to the specific needs of a particular application or service, such as low-latency services for autonomous vehicles or high-bandwidth services for virtual reality experiences \cite{KammounTDAD19, wijethilaka2021survey}. Network slicing offers a multitude of advantages:
\begin{itemize}
\item Flexibility and Agility: Network slices can be provisioned, scaled, and reconfigured to meet dynamic service demands. This allows service providers to quickly adapt to changing market conditions and offer innovative services to their customers.
\item Enhanced Cost-efficiency: Resources are efficiently utilized by dedicating them to specific applications, minimizing wastage. Network operators can optimize their infrastructure investments by allocating resources only where they are truly needed \cite{ChahbarDD21}.
\item Multi-domain and Multi-tenant Support: It empowers the creation of isolated network environments, catering to a variety of users and services across different administrative domains. This enables network operators to offer secure and reliable network slices to multiple tenants, such as enterprises and content providers\cite{DAN21}.
\end{itemize}
In addition, network slicing has the potential to revolutionize various sectors, including:
\begin{itemize}
\item Mobile Network Operators (MNOs): Offering differentiated services with customized SLAs to diverse customer segments \cite{raman2022wireless}. MNOs can leverage network slicing to create high-performance slices for mission-critical applications like industrial automation or remote surgery, while also providing basic connectivity slices for less demanding users.
\item Content Providers: Delivering high-quality streaming experiences with guaranteed bandwidth and low latency. Content providers can create dedicated network slices to ensure smooth and uninterrupted delivery of video and audio content to their subscribers \cite{mehmood2023intent}.
\item Enterprises: Establishing secure and reliable private networks for mission-critical applications. Enterprises can leverage network slicing to create private slices for their internal applications, ensuring isolation, security, and predictable performance \cite{Dandoush_2023}.
\end{itemize}
{\bf Limitations of Current Orchestration and Management}
While network slicing holds immense promise, current orchestration and management approaches primarily rely on SDN based solutions aligned with the specifications of centralized Network Functions Virtualisation (NFV); Management and Orchestration of ETSI\cite{etsisdnfv, etsi} or with a Multi-Level Delegation Architecture as described in \cite{Dandoush_2023}. These approaches face challenges in handling the complexity of new service demands with diverse QoS requirements.
For instance, imagine a scenario where a manufacturing company requires a network slice for its factory automation system. This slice would necessitate ultra-reliable low-latency communication for real-time control of robotic equipment. A management system governed by a central authority (e.g., a network operator) might struggle to efficiently manage such a slice, especially if the factory spans multiple geographical locations with a slice to span different network operators (mutli-administrative domains). In fact, each domain might have its own management system and protocols, making it difficult to achieve seamless slice creation, management, and resource allocation across these boundaries. Moreover, the user will be unable to express the exact and correct requirement and needs of its slice, but rather it will make a high level slice request expressed in a human manner far from a concrete technical request as assumed in the current references \cite{etsi, NESMO, DAN19, DAN21, ChahbarDD21}.

{\bf Motivation: The Need for collaborative Intelligent Management}
The dynamic nature of network slices, coupled with the heterogeneity of services and multi-domain environments, necessitates not only an intelligent but also a collaborative management solution. 
A multi-agent system of Large Language Models (LLMs) can play a pivotal role in addressing challenges and issues in the telecom sector \cite{bariah2023large}. In particular, we believe that they are able to interpret user intents expressed in natural language can facilitate seamless communication between orchestrators across different domains in the slicing context. They are able also to expose an abstracted view of the network topology of a given region/autonomous system to the collaborators (e.g., the other operators or slice providers). The agents distributed across the access, transport, core and cloud networks can collaborate based on the exposed abstracted pieces of information between them in order to standardize and translate the user intents into specific technical requirements, enabling efficient creation or modification of slices that span multiple administrative domains. Furthermore, LLMs contribute to adaptable resource allocation by dynamically analyzing patterns, determining optimal computing, memory, storage, and bandwidth resources for efficient network functioning. In addition, the agents can help in mapping the chain of network functions required for the service requesting the slice to the shared infrastructure across the different parts of the network domains. LLMs also automate complex network configurations and instruction generation, ensuring accuracy and adherence to best practices. In real-time monitoring and analysis, LLMs can report results to network controllers, allowing for proactive measures to address issues promptly and ensure uninterrupted network performance. To the best of our knowledge, this is the first work in the literature to explore the potential of LLMs for Self-Driving Network Slices powered by LLMs.


The rest of this paper is organized as follows: 
Section \ref{sec:llm} introduce the basics of Large Language Models. In Section \ref{sec:propo} the multi-agent LLMs based approach is described from both the user and operator perspectives. The challenges of applying LLMs for network management and orchestration is discussed in Section \ref{sec:challenge} and future directions are presented. Section \ref{sec:con} concludes our work.

\section{Large Language Models Overview}
\label{sec:llm}
In this section we will introduce briefly the Large Language Models (LLMs) and mention some recent use cases. 

LLMs are a type of artificial neural network architecture known as transformers, first introduced by Vaswani et al. in 2017 \cite{Atten17}. Transformers deviate from traditional recurrent neural networks (RNNs) by relying solely on an attention mechanism to capture relationships between elements within the input data. The architecture of a transformer consists of encoder and decoder blocks stacked in series. The encoder block processes the input data (text, code) through multiple layers, with each layer utilizing a so-called ``self-attention'' mechanism. This mechanism allows the model to attend to different parts of the input sequence simultaneously, understanding how each element relates to the others. This is crucial for capturing long-range dependencies within the data \cite{Atten17}. In such a system, the decoder block utilizes the encoded information from the encoder to generate an output. It also employs a self-attention mechanism to ensure the generated output is consistent with the context established in the previous steps. Additionally, the decoder uses an encoder-decoder attention mechanism to attend to relevant parts of the encoded input during generation. During the training of these artificial neural networks, LLMs are exposed to massive datasets of text and code. These datasets can encompass e-books, on-line articles, code repositories, and online conversations. Through a process called backpropagation, the LLM adjusts the weights and connections within its neural network to minimize the difference between its predictions and the actual data. Over time, the LLM develops the ability to recognize patterns, understand complex relationships, and generate human-quality outputs \cite{LLM19}. his empowers them to generate human-quality text, translate languages, create diverse creative content, and answer your questions in an informative way. The LLMs started to be applied in different fields from text and image generation to network log analysis and security threat detection. For example, a recent study in \cite{mani2023} explores the potential of LLMs to generate task-specific code from natural language queries to enhancing network management. In another example, Rawat et al., \cite{Rawat2022} analyse the different use of LLMs including virtual assistants to provide real-time enhanced customer support.


\section{Basic E2E Network Slicing Architecture}
\begin{figure*}[!t]
\centering
\includegraphics[width=\textwidth]{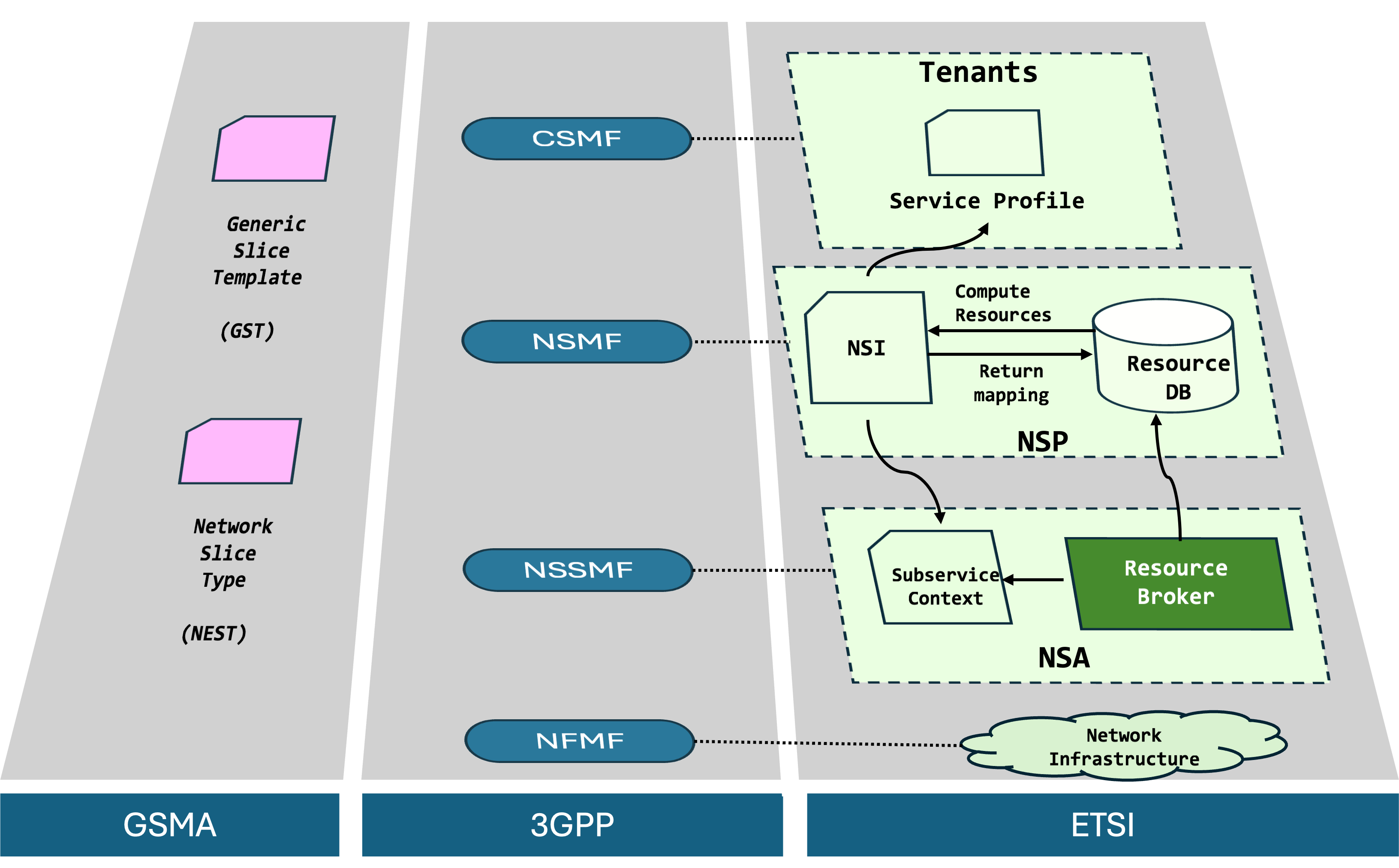}
\caption{SDO's views on Network Slicing - GSMA, 3GPP and ETSI}
\label{fig:slicesdos}
\end{figure*}

Before presenting our approach of integrating LLMs into Network Slicing framework, a general view of the E2E Network Slicing as identified by ETSI NGP workgroup~\cite{etsiNS}(one of the Standards Developing Organizations (SDOs) working on setting the standards of the new generation networks) is introduced in this section. Figure~\ref{fig:slicesdos} illustrates the reference network slicing architecture as comprehensively studied in \cite{DAN21}. This architecture is a simple way to represent slicing from the vision of the different SDOs like 3GPP and ETSI. This will be used hereafter to highlight how LLMs can empower this promising era. In this architecture three actors are collaborating: 

\begin{itemize}
    \item Tenants: Consumers of NS services, which can be NS-as-a-Service (NSaaS).
    \item NS Providers (NSPs): Providers of access to NS instances.
    \item NS Agents (NSAs): Entities with complete view and control over their network domain.
\end{itemize}


The ETSI Network Slicing (NS) architecture encompasses several pivotal operations, each contributing to the seamless orchestration and deployment of end-to-end network slices. These operations play a crucial role in fulfilling tenant requirements, optimizing resource utilization, and ensuring dynamic adaptability to changing network conditions.

\emph{Service Profile Preparation:} The initial step involves a tenant preparing a \emph{service profile} that articulates the desired NS. This profile includes a \emph{service graph} and additional attributes such as service type, profile identifier, and subscribing entity. The service graph delineates nodes in terms of compute, storage, and service instance types, along with slice constraints represented as edges such as the link capacity.

\emph{Resource Database Aggregation:} Within the NS Provider (NSP), the \emph{resource database} (DB) aggregates topological information from various domains in the underlying infrastructure. The resources DB provides a comprehensive view of the abstracted physical infrastructure from multiple Network Slice Agents (NSAs).

\emph{NS Instance Creation:} The NSP creates the \emph{NS Instance (NSI)} based on the received service profile from the tenant. The NSI represents the NSP's view of the NS and involves computations and mappings of the service profile to abstracted infrastructure elements stored in the resources DB.

\emph{NS Delegation and Segment Deployment:} Upon successful computation and mapping, the NSP delegates \emph{segments} to NSAs. A segment represents the set of paths and nodes (e.g., a load balancer function or a firewalling virtual function) allocated to a specific NSA, contributing to the deployment of the overall logical end-to-end NS. The \emph{NS delegation function} ensures efficient interconnection of these segments.

\emph{Tenant-Initiated Resource Augmentation:} Tenants can dynamically scale their allocated resources during runtime using the \emph{subnet augment function}. This operation enables tenants to submit \emph{augment requests}, such as modifying latency constraints, triggering operations for system stability and resource adjustments.


The 3GPP vision of slicing is shown in Figure \ref{fig:slicesdos} which can be seen as complementary to the ETSI architecture \cite{DAN21}). 
Based on the customer's SLA, three main 3GPP functions are involved to manage a network slice. The Customer Service Management Function (CSMF) gathers service requirements from the customer (e.g., the user of a remote surgery service), converts them into NS requirements (a technical Service profile), and forwards them to the NS Management Function (NSMF), which oversees Communication Services provided by the Network Operator in a given administrative domain.

The NSMF handles the creation and management of NS Instances (NSIs) based on translated service profile. The NS subnet management is provided then by the NS Subnet Management Function (NSSMF) at a lower level (the SDN controller level in the ETSI architecture).

NSs and NS Subnets are supplied by the operators owning the infrastructure, where the NSMF and NSSMF are located.
These key operations collectively form the foundation of Network Slicing architecture, offering a comprehensive framework for technology-agnostic network slicing.

\subsection{Limitations and Challenges in Applying Network Slicing Framework}
\label{sec:ns_limitations}

Applying the ETSI/3GPP Network Slicing (NS) framework in scenarios involving multi-administrative domains, such as network segments managed by different operators, and highly heterogeneous services and slice profiles presents a set of challenges and limitations. This section discusses key considerations in addressing these complexities.

\begin{itemize}
    \item \textbf{Interoperability and Standardization:} Achieving seamless interoperability across diverse administrative domains requires robust standardization efforts.

    \item \textbf{Security and Trust:} Managing security across multiple administrative domains with distinct policies is complex. Especially, when dealing with a heterogeneous mix of services crossing an access network (e.g., 5G/6G), an IP backbone network, and a cloud computing resources where the application is deployed.

    \item \textbf{Resource Coordination and Allocation:} Coordinating resources across diverse administrative domains with different policies is challenging due to heterogeneous service requirements and lake of complete view for affecting overall optimization.

    \item \textbf{Policy and Governance:} Establishing a unified policy framework that aligns with various governance models is challenging.

    \item \textbf{Dynamic Service Orchestration:}
Orchestrating highly heterogeneous services in real-time requires dynamic control mechanisms. In fact, complexity in orchestrating services may result in delays and challenges in meeting dynamic service demands.

    \item \textbf{Slice Lifecycle Management:} Diversity in slice profiles may lead to difficulties in managing and adapting to changes in services and resource requirements.

    \item \textbf{Regulatory Compliance:} Ensuring compliance with diverse regulatory frameworks across domains requires continuous monitoring.

    \item \textbf{Service Level Agreements (SLAs):} Defining and adhering to SLAs that satisfy diverse service and profile requirements is complex.
\end{itemize}

Addressing these challenges necessitates collaborative efforts among standardization bodies, operators, and service providers to establish common practices, interfaces, and protocols that facilitate the seamless deployment of network slices in a multi-administrative domain and heterogeneous services environment.

\section{Proposed LLM-based Slicing Management approach}
\label{sec:propo}
\begin{figure*}[!t]
\centering
\includegraphics[width=\textwidth]{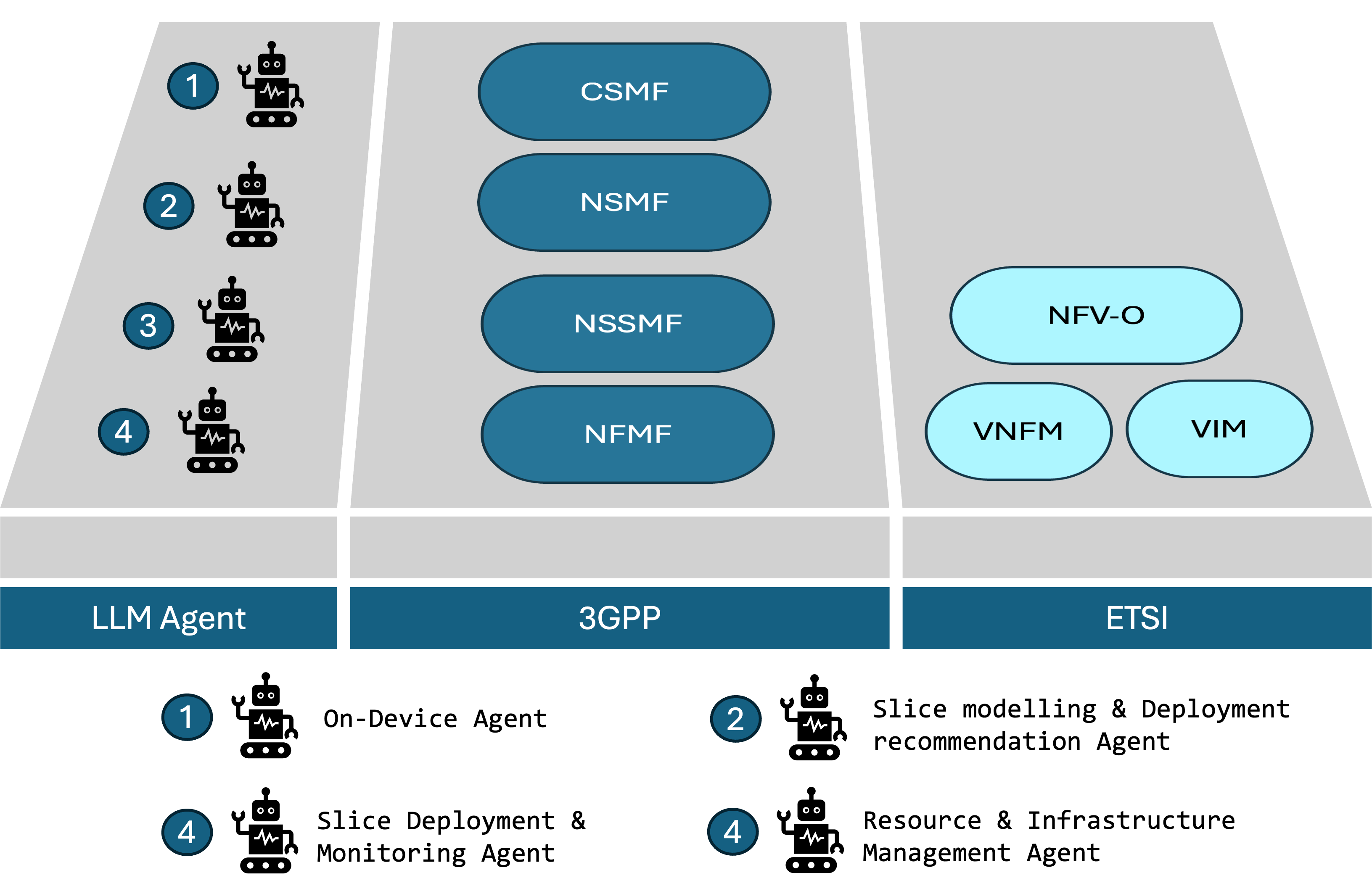}
\caption{LLM Agents scope mapped to 3GPP and ETSI related management functions responsible for network slice deployments}
\label{fig:llmsdos}
\end{figure*}

In this section, we describe our proposed approach for utilizing Multi-Agent Systems (LLMs) as a transformative solution for network slicing management. Figure \ref{fig:llmsdos} describes how the set of LLM agents can be mapped to management frameworks introduced by SDOs such as ETSI and 3GPP. Each agent within the system represents a distinct administrative domain, or alternatively, a Software-Defined Networking (SDN) controller responsible for specific segment subnet configurations and data/log/statistic collection from the underlying infrastructure. Notably, an agent can also function as an SDN orchestrator, synthesizing a high-level abstraction of network, storage, and computing resources obtained from the agents working at the SDN controllers level in each domain. This orchestrator agents plays a pivotal role in exposing and communicating an abstracted view of the topology and the available capacity to other agents (i.e., on the other administrative domains). Through this cooperative and intelligent multi-agent system, we envision the collaborative management of end-to-end network slicing that overcomes the challenges and limitations posed by the deployment of the SDOs Network Slicing frameworks in multi-administrative domains with highly heterogeneous services as discussed in Section \ref{sec:ns_limitations}. 

\begin{figure*}[!t]
\centering
\includegraphics[width=\textwidth]{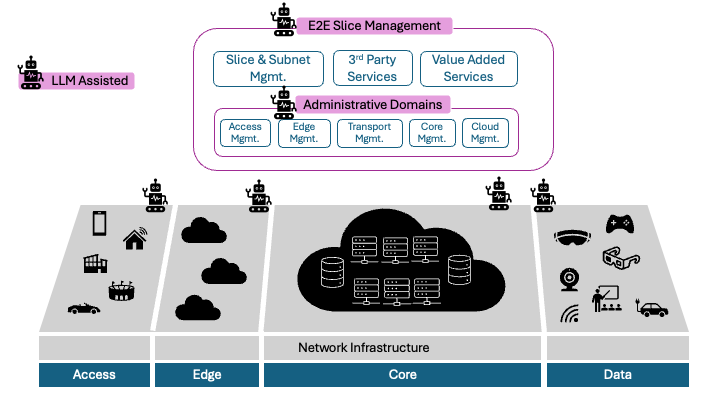}
\caption{High level view of LLMs Assisted Network Slice Management}
\label{fig:e2ellmhlv}
\end{figure*}

The general position of the LLMs multi-agent in E2E Network Slicing Management and Orchestration is illustrated in Figure \ref{fig:e2ellmhlv}. From assisting the user in making his high level intent based slicing request (e.g., Generic Slice Template GST in GSMA reference) to the deployment and monitoring of the different slice functions located at the access network, edge, core and cloud where the applications are hosted. 

LLM Assisted functions can seamlessly collaborate and coordinate with each other thereby offering enriched user experience to both categories of users:  (i) Network Operators and (ii) Network Consumers. In following subsections, we will explain the improved user experience flows for each category of users.

\subsection{From Customer request to auto deployment} 
\subsubsection{Service Consumer point of view}
\begin{figure*}[!t]
\centering
\includegraphics[width=\textwidth]{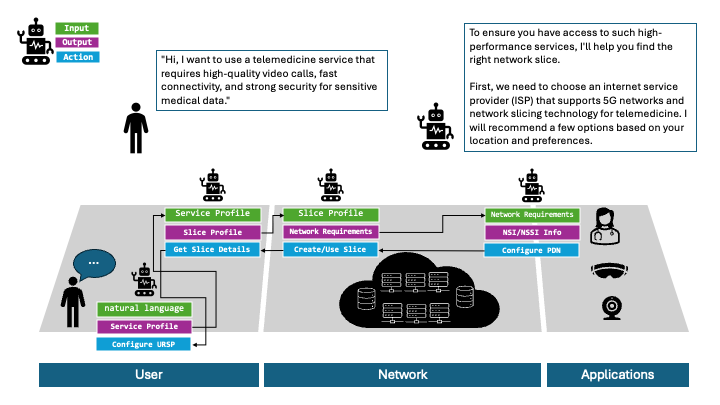}
\caption{LLM Assisted End User Network Slice Intent Translation}
\label{fig:llmuserintent}
\end{figure*}

Large Language Models (LLMs) play a crucial role in simplifying the interaction between end users or applications and the network orchestration system. When a user initiates a network slice request, she/he often communicates her/his needs in natural language, specifying high-level requirements such as good connectivity, low cost or premium service, high or law ``connection speed'', and reliability. A LLM model as shown in Figure \ref{fig:llmuserintent} can translate this high level request to a more concrete service profile, identifying more technical requirements such as latency, throughput, and reliability according to the slice template request standardized by GSMA/3GPP. In other words, LLMs excel in translating these user requests into specific technical requirements that can be understood and executed by the network orchestrator(s). Moreover, if the required constraints cannot be met due to capacity limitation on some portion of the slice, the agent can negotiate (e.g., by sending a specific prompt) with the corresponding agent located at the access network level to negotiate a relaxed version of the requirements that can be allocated and initiated if the user can approve them. Leveraging their natural language understanding capabilities, LLMs ensure precise communication between users and the network slice provider related agents.
Figure \ref{fig:llmuserintent} shows a hypothetical scenario where an end user is requesting one or two slices for Telemedicine service. In fact, based on the scenario, a telemedicine service can consume two slices, one with high throughput for the remote visio and video streaming and another with an ultra low latency requirement used to send commands remotely to a medical devices (e.g., active catheter or robots). The user simply states their intent of consuming a Telemedicine service along with the requirement for high quality video calling and security, considering the privacy and safety concerns related to the health data. In fact one can imagine a service catalogue with innovative services as explained in \cite{Dandoush_2023} to simplify the creation of the corresponding slice or joining an existing one. For every service in the catalogue, a chain of the required functions is already known and predefined. However this list cannot cover all the possible requests a user can make, and can limit innovation. That is why a chatbot agent can assist the user at the early stage to form any request regardless the predefined catalogue idea. The agent then can prompt the corresponding agent at the NS provider to suggest a new chain of functions based on its knowledge and continuous learning of different patterns. The user usually has zero knowledge how the underlying network infrastructure is being built and what network slices are available or need to be created in-order to deliver the Telemedicine or another innovative service. Here a LLM agent that co-reside on the user equipment (a PC or a medical device) can quickly provide on-device NLP experience. The LLM agent can synthesize and understand the user’s intent and efficiently translate that to a 3GPP service profile in terms of standard service attributes that can be exchanged with a standards complaint Communication Service Management Function (CSMF) and Network Slice Management Function (NSMF) respectively.

Here in this case, multiple LLM agents can work at different administrative domain to execute a control task. For example LLM agents at NSMF can coordinate with LLM agents as Network Slice Subnet Management functions such as Core, Access and Transport subdomains to effectively configure underlying network infrastructure as per the network intents specified in the Slice profile. In fact, intelligent mapping of one or more network slice profiles to a service profile can be efficiently performed a utility LLM agent that is sitting between NSMF and NSSMFs respectively.

\begin{figure*}[!t]
\centering
\includegraphics[width=\textwidth]{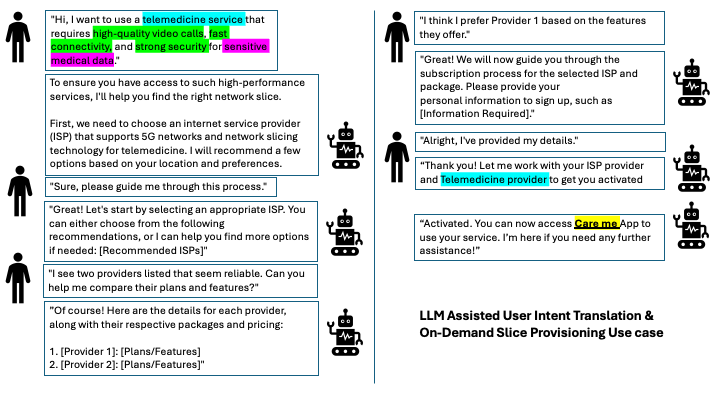}
\caption{A hypothetical interaction between an end user and an intelligent LLM, negotiating over a telemedicine service}
\label{fig:llmuserintenthypo}
\end{figure*}

Introducing LLM assisted capability in service fulfillment will not only improve automation and time to  efficiency, but also will vastly improve end user experience. Figure \ref{fig:llmuserintenthypo} depicts a hypothetical natural language conversation between an end user and the on-device LLM agent. The conversation starts with the user stating their intent to consume a telemedicine service. The LLM agent walk-through all the steps needed to successfully sign-up the user to the telemedicine service while abstracting away all the complexities in network service provisioning flow.

LLM agent first extracts the keywords from the user’s intent (such as telemedicine service, high quality video calls, security, sensitive medical data) and deduces that for this service to be provided, a special network slice has to be utilized. LLM Agent also deduces that a 5G NR access or a fiber optical broadband access is very crucial for the ultra low latency service requirement.

\subsubsection{Network Operator point of view}

The potential for LLMs to assist Network Operators in planning and deployment of network slices is enormous. In this section, we explore some of the possible use-cases. As in the case of network service consumer, a network operator user can express their intent of deploying one or more network slices along with specific optimization goals for each of the network slice. This intent along with information about the underlying network infrastructure (such as access, transport, core and datacenter network pieces of information) can be used by series of LLM agents to coordinate a network modelling exercise.

\begin{figure*}[!t]
\centering
\includegraphics[width=\textwidth]{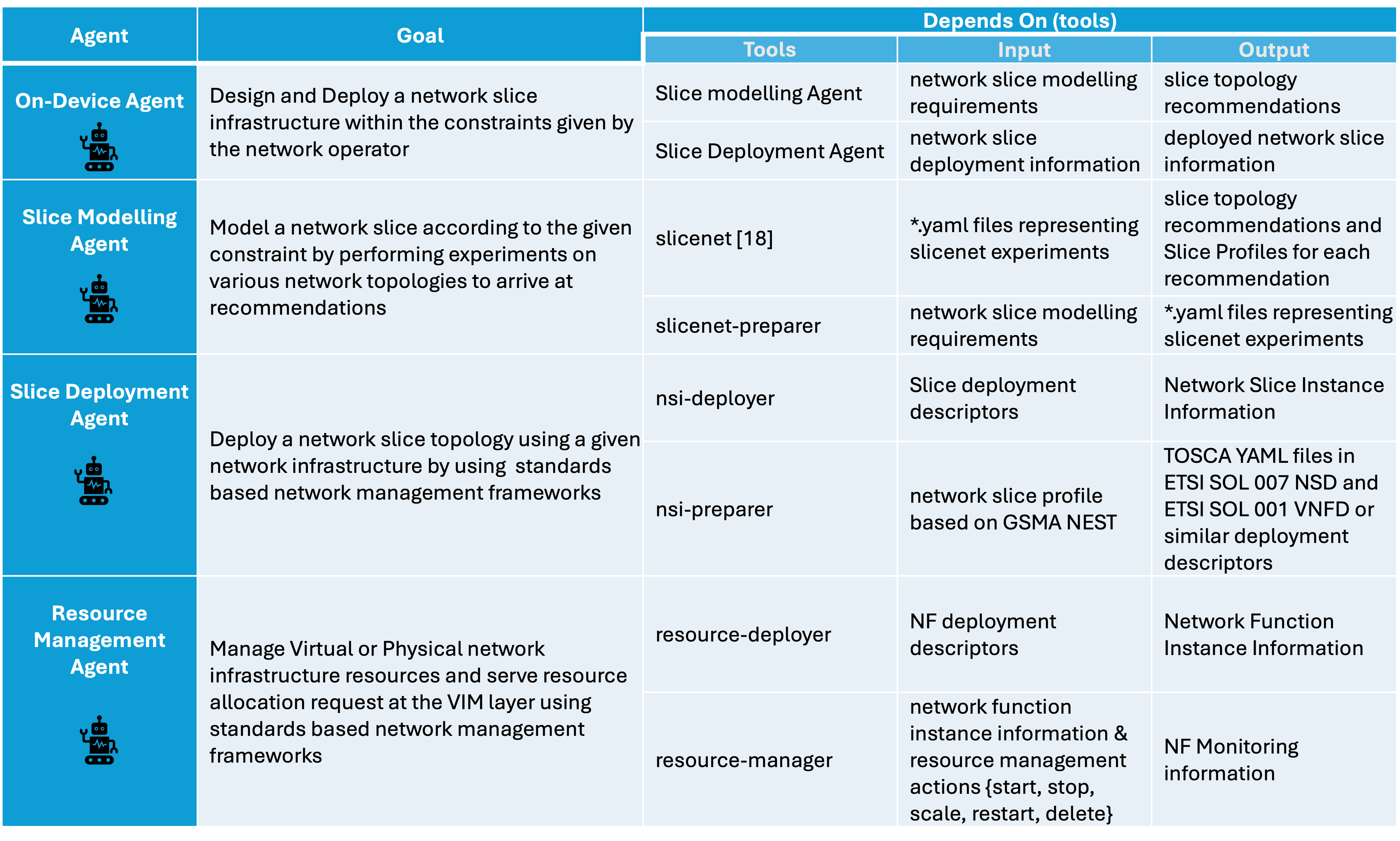}
\caption{Slice LLM Agents}
\label{fig:slicellmagentsdefs}
\end{figure*}

Figure \ref{fig:slicellmagentsdefs} describes the composition of multi-agent framework for LLM assisted network slice design and deployment, which can be used by a network operator. Each agent is created with a defined role, goal, inputs to consume and outputs to produce. Also each agent is equipped with a set of tools to carry out their respective tasks. The \textit{On-Device Agent} is the higher-level user facing agent that resides within the user equipment to facilitate the user interaction with the Network Slice Provider (NSP) or the Network Operator (NOP). This agent has access to  \textit{Slice modelling Agent} and \textit{Slice Deployment Agent} as part of \textit{tools} declaration within the agent's definition. Each tool has its own \textit{inputs} and \textit{outputs} respectively. LLM learns what input to pass and what output to collect based on the respective tool's definition. A complete sample listing of network slice LLM agent definition is provided in the appendix \ref{appendix:agentListing} section as a Yaml \footnote{YAML is a human-friendly data serialization language for all programming languages.} template.

\begin{figure*}[!t]
\centering
\includegraphics[width=\textwidth]{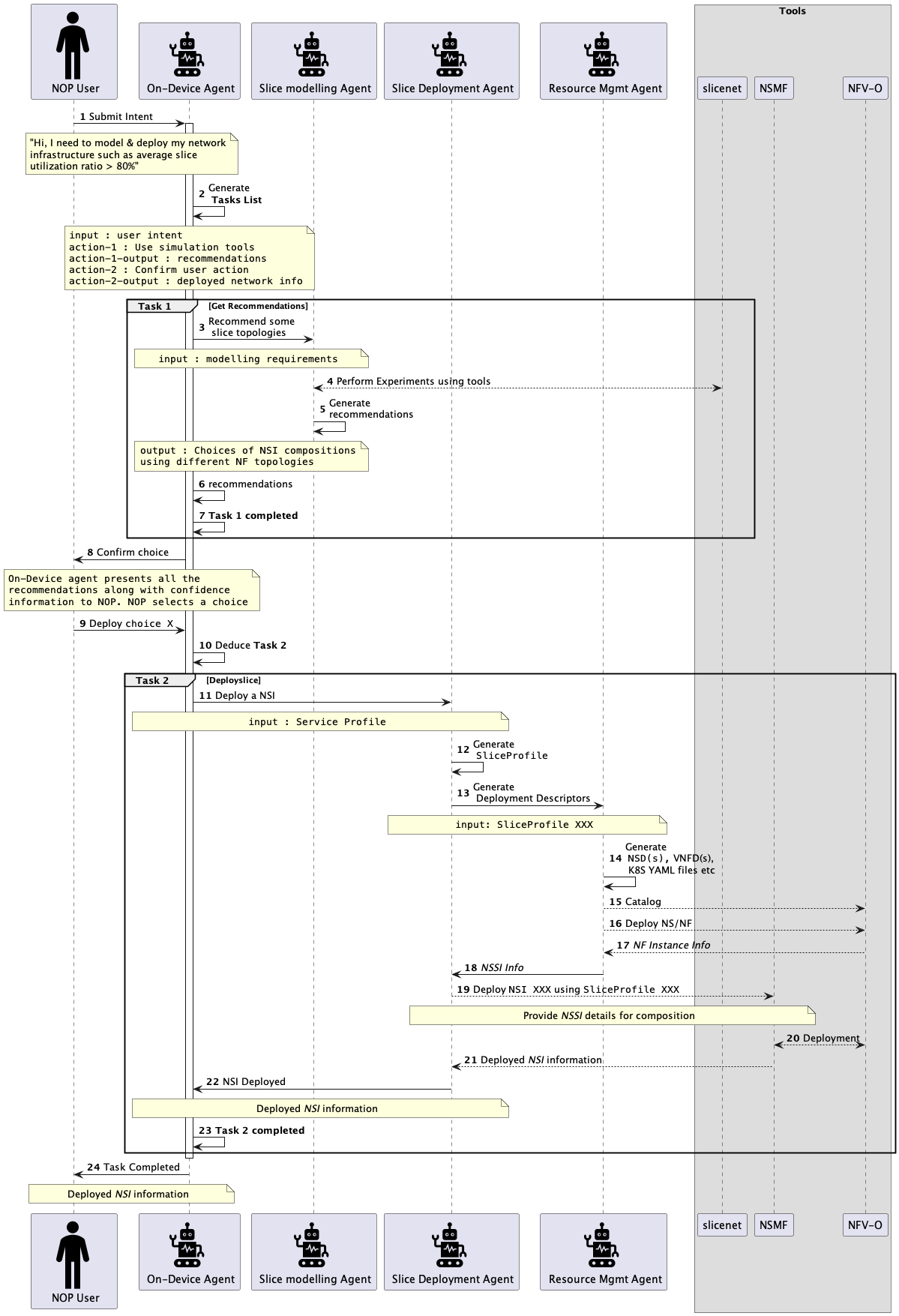}
\caption{Sequence of interaction among Slice agents to design and deploy a network slice against a specific optimization constraint}
\label{fig:nopsliceagentseq}
\end{figure*}

Figure \ref{fig:nopsliceagentseq} explains how these LLM agents can work collaboratively for a network operator use-case. Consider a scenario where a network operator would like to design and deploy a network slice instance within an underlying network infrastructure with limited resources such that, the average slice utilization ratio is greater than 80 percent (for optimizing power and cost). Without LLM assistance, this process would involve considerable effort as a network planning activity involving a long journey. However this entire experience can be rapidly automated with LLMs in picture.

In this typical use-case, the network operator simply expresses its intent to the \textit{On-Device} agent which is capable of using \textit{Slice modelling} and \textit{Slice Deployment Agents}. First it deduces the tasks to be followed. In this case, it learns first the slice modelling has to be done to generate some recommendations on different slice topologies that would fit the optimization constraint of maintaining overall slice utilization ratio of greater than 80 percent. In-order to perform this task, it chooses the \textit{Slice modelling Agent} based on the description. Since this tool requires some input, \textit{On-Device agent} generates that input based on its knowledge and call the \textit{Slice modelling Agent} to perform \textbf{Task 1} as shown in Figure \ref{fig:nopsliceagentseq}. \textit{Slice modelling} Agent in-turn uses another tool like slicenet \cite{slicenet} to perform the necessary experiments and modelling of the network to fit the optimization criteria. Based on the insights, \textit{Slice modelling} agent further generates some recommendations on a different Network Slice Instance (NSI) topologies that could fit the optimization goal and pass this as the output to \textit{On-Device} agent (refer to step \#7 in sequence diagram depicted in Figure \ref{fig:nopsliceagentseq}. This is further presented back to the NOP user to confirm its choice (as shown in step\#8 in the figure). Here the user has option to either accept one of the choice provided by the LLM or provide reinforcement learning from human/AI (RLHF) feedback to fine tune the modelling task as needed. If the NOP user proceeds to select a choice, \textit{On-Device} agent proceeds to \textbf{}{Task 2}. In this task, \textit{On-Device} agent works with \textit{Slice Deployment and Resource Management Agents} respectively. These lower level agents are capable of generating standards based deployment descriptors such as \textit{SliceProfile\cite{28541}, Network Service Descriptor \cite{sol007}, Network Function Descriptor \cite{sol001}} and also interact with Network Slice Management Systems, NFV-Os to deploy the network slice instances. Post deployment, the information about the created NSI(s) along with other site specific information is relayed back to the \textit{On-Device} agent thereby concluding \textbf{Task 2}.

Last, concerning the deployment of Task2, by LLMs assisting in the generation of specific requirements, such as computing, memory, storage, and bandwidth resources accross each administrative domain,  the NSP ensures that the orchestrators at different domains are equipped with accurate and standardized information templates, promoting consistency and interoperability (e.g., using TOSCA  Yaml as shown in Appendix \ref{appendix:agentListing}). Moreover, LLMs support the mapping of user requirements to network functions and services. This involves analyzing the placement of functions, which is a critical step in optimizing resource allocation and then converting the map into actionable configurations, promoting a harmonized approach to network slicing management.
 Also, one of the critical challenges in network slicing is the optimal allocation of resources to different slices. LLMs, with their ability to analyze and understand complex patterns and requirements, can determine dynamically the appropriate computing, memory, storage, and bandwidth resources (e.g., auto-scaling up or down) at a given time based on the collected statistics from the southbound interfaces such as P4 or OpenFlow, resulting in a more efficient network functioning. This adaptability is essential for accommodating the varying needs of different applications and services and increases NSP profit while keep respecting the SLA of the custemers.

\section{Challenges and Future Directions}
\label{sec:challenge}
\subsection{Current Challenges}

Integrating Large Language Models (LLMs) into end-to-end network slice management presents several challenges and requires careful consideration:

\begin{enumerate}
    \item \textbf{Data Set Availability:} Acquiring relevant and diverse datasets for training and fine-tuning LLMs within a network management context remains a challenge. Simulation techniques can partially address this issue by generating synthetic data, but ensuring the authenticity and representativeness of such data is crucial \cite{slicenet}.
    
    \item \textbf{Computing Power and Hosting:} The computational demands of LLMs, especially large models, can strain existing infrastructure. Advances in developing lighter versions of LLMs such as Phi-1 \cite{gunasekar2023textbooks}, optimized for deployment on edge devices or cloud environments, are essential for scalability and cost-effectiveness.
    
    \item \textbf{Network Security Backdoors:} LLMs, like any complex software system, can potentially have vulnerabilities that malicious actors may exploit as backdoors \cite{ ya2024}. Utilizing LLMs themselves for continuous analysis and mitigation of such security risks is a proactive approach and an active research area.
    
    \item \textbf{Trust and Validation:} Ensuring the trustworthiness and reliability of LLM-assisted processes is paramount. Legal and regulatory compliance, such as avoiding SLA violations and network neutrality violation, requires rigorous validation and adherence to standards mainly with the difficulty of fine-tuning the LLM models with the network data due to the unexciting of large deployed slicing technologies. Testbeds and Simulation techniques play a crucial role in testing and validating LLM integration while addressing trust-related concerns. Collaboration with communication regulatory authority and operators is vital for the validation phase before to go to commercialization. 
    
    \item \textbf{Interoperability and Standards:} Integrating LLM-assisted functions across heterogeneous network environments requires adherence to interoperability standards. Developing common protocols and interfaces for communication (e.g., northbound interfaces), prompts and data exchange among LLMs and network components is essential for seamless integration.
    
    \item \textbf{Privacy and Data Protection:} LLMs often operate on sensitive data, raising concerns about privacy and data protection  \cite{wu2024secgpt}. Implementing robust encryption methods, data anonymization techniques, and access control mechanisms can mitigate privacy risks associated with LLM usage.
    
\end{enumerate}

\subsection{Future Directions and Solutions}

To address these challenges and leverage the potential of LLMs for end-to-end network slice management, several future directions and solutions can be explored:

\begin{enumerate}
    \item \textbf{Collaborative Learning:} Implementing collaborative learning frameworks where multiple LLMs collaborate and share knowledge can enhance their collective intelligence and decision-making capabilities \cite{bariah2023large}.
    
    \item \textbf{Federated Learning:} Utilizing federated learning techniques to train LLMs across distributed network environments while preserving data privacy and security \cite{ye2024openfedllm}.
    
    \item \textbf{Explainable AI:} Enhancing the explainability of LLMs to provide transparent insights into their decision-making processes, fostering trust and understanding among stakeholders \cite{wu2024usable}.
    
    \item \textbf{Continuous Monitoring and Auditing:} Implementing robust monitoring and auditing mechanisms to detect and mitigate potential biases, errors, or security vulnerabilities in LLM-assisted processes.
    
    \item \textbf{Ethical Considerations:} Incorporating ethical frameworks and guidelines into the development and deployment of LLMs to ensure responsible and fair use of AI technologies in network management.
    
    \item \textbf{User-Centric Design:} Designing LLM-assisted interfaces and workflows with a focus on user experience and usability, enabling intuitive interaction and effective utilization of LLM capabilities.
    
    \item \textbf{Industrial, academic and governmental Collaboration:} Promoting collaboration and knowledge-sharing among industry stakeholders, academia, and regulatory bodies to address common challenges and foster innovation in LLM-enabled network management.
\end{enumerate}

By addressing these challenges and exploring innovative solutions, the integration of LLM multi-agent systems into end-to-end network slice management can unlock significant benefits in terms of efficiency, automation, and optimization of telecommunications networks.

\section{Conclusion}
\label{sec:con}
In the era of multi-domain network slicing, the integration of large language models (LLMs) and advanced AI technologies redefines network management paradigms. By embracing intelligent, context-aware orchestration, organizations can build responsive, secure, and efficient network infrastructures spanning diverse administrative domains. This research illuminates the path forward, where AI-driven multi-domain network slicing becomes the cornerstone of next-generation networks, ensuring unparalleled connectivity and user satisfaction. In this context, we proposed a framework that maps agents to the ETSI and 3GPP vision and recent management works of E2E slicing. This framework explains how LLMs can collaborate from assisting the user in initiating the request to deploying, continually monitoring resource consumption, and assisting in auto-scaling and allocation optimization.

However, it's essential to acknowledge the challenges associated with integrating LLMs for end-to-end network slicing, including data set availability, computing power, network security backdoors, trust and validation, interoperability and privacy. Addressing these challenges will pave the way for future research and development efforts in this domain. It is important to explore new collaborative learning techniques, improving explainability of AI decisions, and promoting industry and regulation authority collaboration to establish standards and best practices.

\bibliographystyle{IEEEtran}
\bibliography{ref.bib}

\clearpage
\appendix

\section{Sample LLM Agents Definition with agent role, description and associated tools used by each agent.}\label{appendix:agentListing}

\begin{lstlisting}[language=bash, caption=LLM Assisted Slice Agents Definitions in Yaml template]
llm-assisted-slice-agents:
  - name : 'On-Device Agent',
    goal : 'Design and Deploy a network slice infrastructure within the constraints given by the network operator',
    prompt : 'You are a helpful on-device assistant. You are allowed use the agents mentioned under tools section to accomplish the goal. You should break down the main goal into different tasks and for each task, use the appropriate agent to accomplish that task. For each task, provide the necessary inputs. At the end of each task, collect the output and process it if necessary and chain that output to subsequent tasks if needed.'
    tools :
      - name : 'modeller',
        agent : 'Slice modelling Agent',
        description : 'Use this tool to perform simulation of different network scenarios and collect the recommendations based on experiment insights',
        input : 'network slice modelling requirements',
        output : 'slice topology recommendations'
      - name : 'deployer',
        agent : 'Slice Deployment Agent',
        description : 'Use this tool to perform deployment of a specific network topology and collect the network access information',
        input : 'network slice deployment information',
        output : 'deployed network slice information'
  - name : 'Slice Modelling Agent',
    goal : 'Model a network slice according to the given constraint by performing experiments on various network topologies to arrive at recommendations',
    prompt : 'You are a network slice simulation expert. You are allowed use the agents mentioned under tools section to accomplish the goal. You should break down the main goal into different tasks and for each task, use the appropriate agent to accomplish that task. For each task, provide the necessary inputs. At the end of each task, collect the output, \
    process it if necessary and chain that output to subsequent tasks if needed.'
    tools :
      - name : 'slicenet',
        description : 'Use this tool to perform simulation of different network scenarios and collect the recommendations based on experiment insights',
        input : '*.yaml files representing slicenet experiments',
        output : 'slice topology recommendations and Slice Profiles for each recommendation'
      - name : 'slicenet-preparer',
        description : 'Use this tool to generate slicent yaml files describing the experiment goal, network constraints and network infrastructure information',
        input : 'network slice modelling requirements',
        output : '*.yaml files representing slicenet experiments'  
  - name : 'Slice Deployment Agent',
    goal : 'Deploy a network slice topology using a given network infrastructure by using standards based network management frameworks',
    prompt : 'You are a network slice management expert. You are allowed use the agents mentioned under tools section to accomplish the goal. You should break down the main goal into different tasks and for each task, use the appropriate agent to accomplish that task. For each task, provide the necessary inputs. At the end of each task, collect the output, process it if necessary and chain that output to subsequent tasks if needed.'
    tools :
      - name : 'nsi-deployer',
        description : 'Use this tool to deploy or discover NSI using NSMF or similar orchestrators  and collect network slice instance information including Slice ID and other deployment information',
        input : 'Slice deployment descriptors',
        output : 'Network Slice Instance Information'
      - name : 'nsi-preparer',
        description : 'Use this tool to generate network slice & subnet templates that describes the network slice instance topology based on slice profile',
        input : 'network slice profile based on GSMA NEST',
        output : 'TOSCA YAML files in ETSI SOL 007 NSD and ETSI SOL 001 VNFD or similar deployment descriptors'  
  - name : 'Resource Management Agent',
    goal : 'Manage Virtual or Physical network infrastructure resources and serve resource allocation request at the VIM layer using standards based network management frameworks'
    prompt : 'You are a network infrastructure expert. You are allowed use the agents mentioned under tools section to accomplish the goal. You should break down the main goal into different tasks and for each task, use the appropriate agent to accomplish that task. For each task, provide the necessary inputs. At the end of each task, collect the output, process it if necessary and chain that output to subsequent tasks if needed.'
    tools :
      - name : 'resource-deployer',
        description : 'Use this tool to deploy network function and associated dependencies  and collect network slice instance information including Slice ID and other deployment information',
        input : 'NF deployment descriptors',
        output : 'Network Function Instance Information'
      - name : 'resource-manager',
        description : 'Use this tool to perform LCM and Resource management of network functions',
        input : 'network function instance information & resource management actions {start, stop, scale, restart, delete} ',
        output : 'NF Monitoring information'  
\end{lstlisting}

\end{document}